\documentclass[trackchanges]{aastex701}

\begin{document}

\title{Compressible Turbulence as a Source of Particle Beams and Ion Bernstein Waves in Collisionless Plasmas}

\author[0000-0001-7205-2449]{Chuanpeng Hou}
\altaffiliation{Humboldt Research Fellow.}
\affiliation{Institut für Physik und Astronomie, Universität Potsdam, D-14476 Potsdam, Germany}
\email{}  

\author[0000-0003-2560-8066]{Huirong Yan}
\affiliation{Deutsches Elektronen Synchrotron (DESY), Platanenallee 6, D-15738 Zeuthen, Germany}
\affiliation{Institut für Physik und Astronomie, Universität Potsdam, D-14476 Potsdam, Germany}
\email[show]{huirong.yan@desy.de} 

\author[0000-0003-4268-7763]{Siqi Zhao}
\affiliation{Institut für Physik und Astronomie, Universität Potsdam, D-14476 Potsdam, Germany}
\affiliation{Deutsches Elektronen Synchrotron (DESY), Platanenallee 6, D-15738 Zeuthen, Germany}
\email{}

\correspondingauthor{Huirong Yan}

\begin{abstract}
Unraveling the origin of proton beams and ion Bernstein waves is important to understanding kinetic dissipation in the solar wind. Here we focus on their generation mechanisms, rather than their well-studied roles in instabilities and particle heating. We investigate their formation in collisionless plasmas using high-resolution particle-in-cell simulations of compressible turbulence. At magnetohydrodynamic (MHD) scales, compressive fluctuations are damped via transit-time damping (TTD), naturally producing suprathermal electrons and proton beams. At sub-ion scales, quasi-perpendicular fast modes excite multiple branches of ion Bernstein waves, whose properties agree with predictions from the plasma dispersion relation solver. Under solar wind conditions, TTD remains efficient and provides a natural explanation for the super-Alfvénic proton beams measured in situ. Our results demonstrate that compressive fluctuations play a central role in driving cross-scale energy transfer and kinetic dissipation in collisionless plasma turbulence.
\end{abstract}

\section{Introduction}

During the expansion of coronal plasma into the turbulent solar wind, the temperature decreases more slowly than predicted by adiabatic models \citep{freeman1985cold,hundhausen2012coronal,dakeyo2022statistical}. This long-standing solar wind heating problem has been a central topic in space physics \citep{tu1997two,shi2022acceleration}. In-situ measurements provide a direct means to investigate the extra thermal energy through particle velocity distributions \citep{he2015evidence}. Observations show that proton velocity distributions frequently deviate from a Maxwellian form and exhibit a core–beam distribution, with a beam population drifting along the magnetic field \citep{marsch1982solar,vdurovcova2019evolution,huang2020proton,verniero2022strong}. In the weakly collisional solar wind, these proton beams contain substantial free energy, can excite ion-scale waves, and contribute to solar wind heating through wave–particle interactions \citep{xie2004multiple,jian2009ion,verscharen2013parallel,he2015evidence,verscharen2015deceleration,chen2016multi,bowen2022situ}. As such, proton beams are widely regarded as an important source for non-adiabatic heating in the solar wind.

Due to their importance, the physical properties and formation mechanisms of proton beams have been extensively studied. Observations show that the relative drift speed between the proton core and beam often exceeds the local Alfv\'en speed ($v_{\text{A}}$) \citep{alterman2018comparison} and increases with plasma $\beta$ \citep{tu2004dependence,de2023innovative}, where $\beta$ is defined as the ratio of thermal to magnetic pressure. Near-Sun measurements by Parker Solar Probe further indicate that the average beam speed can reach about 2.5$v_{\text{A}}$ \citep{verniero2022strong}. A variety of mechanisms have been proposed to explain the origin of proton beams, including Coulomb collisions and magnetic-field inhomogeneity \citep{livi1987generation}, cyclotron-resonance-induced diffusion \citep{tu2002formation}, magnetic reconnection \citep{lavraud2021magnetic}, and parametric instability of Alfv\'en waves \citep{araneda2008proton}. However, no consensus has yet been reached, highlighting the need for a unified physical picture linking proton beams to solar wind turbulence.

Beyond proton beams, kinetic waves represent another key feature of the solar wind whose origin remains under active investigation. In addition to ion-acoustic waves \citep{kellogg2024heating}, whistler waves \citep{tong2019statistical,yao2021low}, ion cyclotron waves \citep{jian2014electromagnetic}, and kinetic Alfv\'en waves \citep{salem2012identification,huang2020kinetic}, high-resolution observations from Cluster and MMS have recently revealed that ion Bernstein waves (IBWs), which propagate mainly perpendicular to the magnetic field, also exist in both the solar wind \citep{joyce2012observation,roberts2020possible} and the magnetosheath \citep{turc2023transmission}. IBWs can heat ions through cyclotron resonance and may interact with kinetic Alfv\'en waves through mode conversion \citep{podesta2012need}, thereby providing additional pathways for energy transfer. Similar to proton beams, the physical origin of IBWs is still under debate. One possibility is that IBWs emerge from a turbulent cascade from large scales down to the sub-ion scales \citep{roberts2020possible}. Another possibility is that they are driven locally by non-thermal ion distributions such as ion rings \citep{joyce2012observation,roberts2020possible}.

To understand the origin of these microscopic phenomena, it is necessary to distinguish the contributions of compressible turbulence components \citep{makwana2020properties, zhang2020identification,zhao2024small, pavaskar2024diagnostics, zhang2025machine}. 
Wave analyses based on Parker Solar Probe (PSP) observations have shown that compressible fast magnetosonic modes carry a substantial fraction of the turbulent energy in solar wind turbulence, with their relative contribution increasing toward the Sun \citep{shi2021alfvenic, Zhao2021, zhu2020wave}. These findings indicate that compressible fluctuations play a key role in energy transfer and dissipation in the solar wind \citep{he2015proton, fu2022nature}. Besides, both theoretical and observational advances have demonstrated that compressibility can significantly modify the turbulent cascade \citep{carbone2009scaling,banerjee2016scaling}. In particular, compressible effects can enhance the energy cascade rate and extend the inertial range, implying a more efficient transfer of energy toward kinetic scales \citep{hadid2017energy}. From a theoretical perspective, fast waves can dissipate energy efficiently at magnetohydrodynamic (MHD) scales through transit-time damping (TTD) \citep{yan2004cosmic,zhao2024small,Zhao2026}. At low $\beta$ environment, the turbulent cascade rate of fast waves could exceed the angle-dependent damping rate, allowing them to maintain a continuous cascade into kinetic scales before being fully dissipated \citep{yan2004cosmic,hou2025energy}. In addition to this cascade process, the phase steepening of fast waves may provide a more direct pathway for the generation of small-scale waves due to dispersive effect \citep{suzuki2007cascading,balogh2013physics}.

Motivated by these considerations, this study investigates the cross-scale evolution of compressible turbulence from MHD to kinetic scales. We use fully kinetic particle-in-cell (PIC) simulations to construct a self-consistent turbulence system initialized with fast magnetosonic waves. The results show that both proton beams and IBWs can be generated naturally during the evolution of fast wave turbulence. This finding provides new physical insight into the origins of both beams and IBWs and demonstrates the important role of compressible turbulence.

\section{Simulation Setup}\label{simulation_setup}

Our high spatial and temporal resolution compressible turbulence simulation is performed using the VPIC code \citep{bowers20080,bowers2008ultrahigh,bowers2009advances}. This decaying compressible turbulence is initialized by superposing fast waves with wave vectors $[\pm k_{0x}, \pm 2k_{0x}, \pm 3k_{0x}]$ and $[\pm k_{0y}, \pm 2k_{0y}, \pm 3k_{0y}]$ along the $x$ and $y$ directions, respectively, where $k_{0x}=k_{0y} = 2\pi / L_0$ and $L_0$ is the box size. The use of integer wave vectors avoids artificial field and density discontinuities at the domain boundaries. To capture turbulence at MHD scales, the simulations employ a large computational domain with periodic boundary conditions. The domain size is $256d_i\times256d_i$, resolved by $4096^2$ cells ($\Delta x=0.0625d_i$), where $d_\mathrm{i} = v_\mathrm{A}/\Omega_\mathrm{p}$ is the proton inertial length, $v_\mathrm{A}$ is the Alfv\'en speed, and $\Omega_\mathrm{p}$ is the proton gyrofrequency. Both protons and electrons are treated kinetically with a reduced mass ratio $m_p/m_e=10$, using 500 particles per species per cell. Particles are initialized in three-dimensional velocity space with a total plasma $\beta = 0.5$ and equal proton and electron temperatures ($T_\mathrm{p} = T_\mathrm{e}$). The time step $\Delta t=0.00625\Omega_p^{-1}=0.0625\Omega_e^{-1}$ ensures adequate resolution of electron dynamics. The initial electromagnetic fields and particle moments follow fast-wave polarization relations, with random phases and equal amplitudes, yielding $\delta v_{\rm rms}/v_A\simeq0.23$, where $\delta v_{\mathrm{rms}} = \sqrt{\langle |\delta v|^2 \rangle}$ denotes the root-mean-square of velocity fluctuation.  The nonlinear timescale for turbulence evolution is $\tau_{\mathrm{nl}} \approx 59\Omega_{\mathrm{p}}^{-1} \approx 0.23\tau_A$, where $\tau_A= L_0/v_A$.

\section{Results and Discussion}

\subsection{Angle-dependent turbulence cascade and damping}

\begin{figure*}[ht]
\centering
\includegraphics[width=0.8\textwidth]{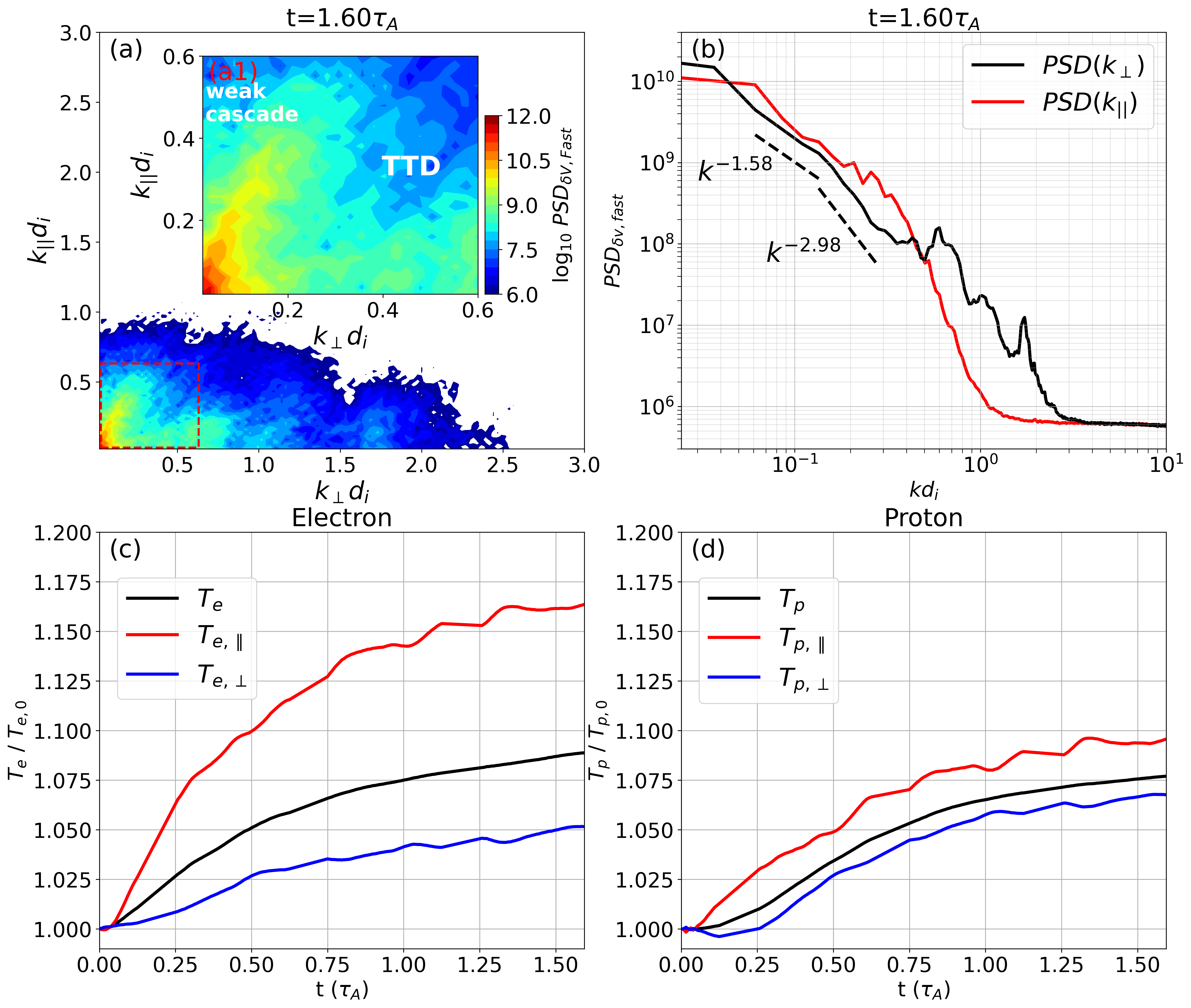}
\caption{Energy distribution of compressible turbulence and the evolution of particle temperatures.
(a) Energy distribution of compressible turbulence in the $k_{\perp}$–$k_{\parallel}$ space. (a1) Zoomed-in view corresponding to the red dashed box in panel (a). (b) 1D integrated PSD. (c) Temporal evolution of the electron temperature components, normalized by initial temperature. (d) The same as panel (c) but for protons.}\label{fig1} 
\end{figure*}

The simulations reveal that fast wave turbulence develops a clear angle-dependent power distribution in wavenumber space (k-space) at $t=1.6\tau_A\approx7\tau_{nl}$ (Figure \ref{fig1}(a)), by which time the power spectral density (PSD) spectrum (Figure \ref{fig1}b) has reached a steady state.  This behavior arises from two main factors. First, the energy transfer proceeds mainly along the radial direction, while coupling between different propagation angles is weak. When the fast wave propagates nearly parallel to the background magnetic field, the associated density and magnetic-field perturbations become very small, and the wave behaves similarly to an incompressible Alfv\'en wave. The cascade rate therefore reaches its minimum in the quasi-parallel direction \citep{galtier2023fast,hou2025energy}. As a result, the fluctuation energy in 2D k-space is weaker near the parallel direction than near the perpendicular direction. Second, the fluctuation energy drops sharply at propagation angles between $40^\circ\sim65^\circ$ (Figure \ref{fig1}(a1)). This reduction is caused by TTD with electrons and protons \citep{zhao2024small,hou2025energy}. The TTD rate can be written as \citep{yan2004cosmic} 
\begin{equation}\label{eq_damping}
   \Gamma ={} \frac{\sqrt{\pi\beta}}{4}\frac{\sin^2\theta_{kB}}{\cos\theta_{kB}} 
\left[\sqrt{\frac{m_e}{m_p}}
\exp\!\left(-\frac{m_e}{m_p\beta\cos^2\theta_{kB}}\right)
+ 5 \exp\!\left(-\frac{1}{\beta\cos^2\theta_{kB}}\right)\right]k v_f,
\end{equation}
where $\theta_{kB}$ is the propagation angle relative to the magnetic field, and $v_f$ is the fast magnetosonic speed.

\begin{figure*}[ht]
\centering
\includegraphics[width=0.5\textwidth]{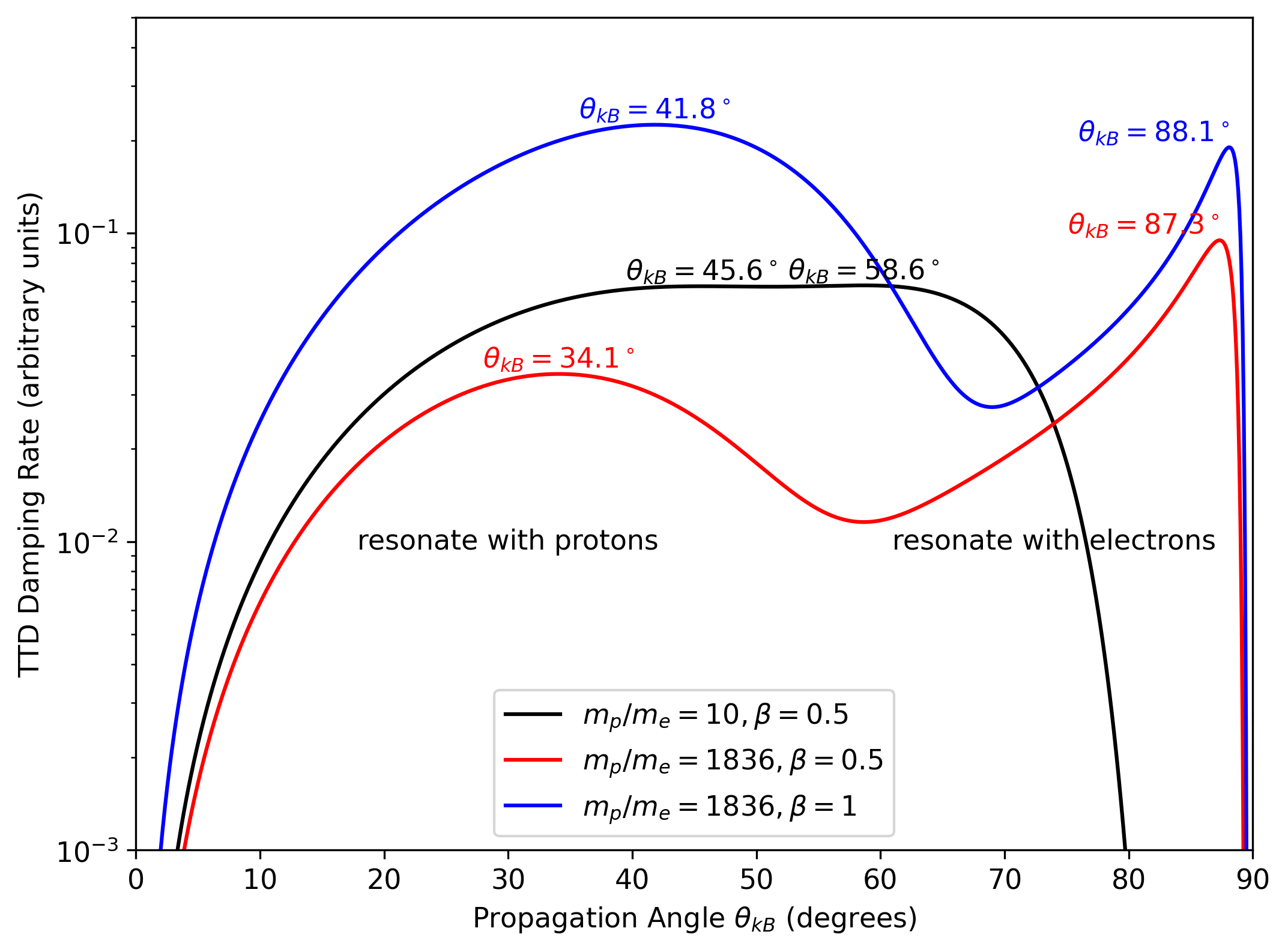}
\caption{TTD damping rate calculated from Equation~(\ref{eq_damping}) as a function of propagation angle $\theta_{kB}$ for different proton–electron mass ratios $m_p/m_e$ and plasma $\beta$. We mark the $\theta_{kB}$ at which the TTD rate peaks.}\label{fig2} 
\end{figure*}

Equation~\ref{eq_damping} is valid in low-$\beta$ turbulent environments, as demonstrated in \cite{hou2025energy} using PIC simulations with relatively larger injection scales. It also remains applicable for $\beta \sim 1$ \citep{ginzburg1962propagation}, as it reproduces an angular dependence of the damping rate $\Gamma$ that is consistent with results obtained from WHAMP \citep{Zhao2026}, which solves the general wave dispersion relations in plasmas. Equation \ref{eq_damping} depends on both $\beta$ and $m_p/m_e$. For a realistic $m_p/m_e \approx 1836$ and $\beta = 0.5$, the damping rate exhibits two distinct peaks: one associated with proton resonance with fast waves at $\theta_{kB} \sim 34^\circ$, and the other associated with electron resonance at nearly perpendicular propagation, $\theta_{kB} \sim 87^\circ$. These peaks arise from the resonance condition $v_{\parallel} = \omega/k_{\parallel} = v_f/\cos\theta_{kB}$. In contrast, when a reduced $m_p/m_e = 10$ is used in these simulations, the proton and electron resonance peaks merge into a single broad angular range of $\theta_{kB} \sim 40^\circ$--$65^\circ$ (Figure~\ref{fig1}(a1) and Figure~\ref{fig2}).

The effect of TTD is more clearly seen in the one-dimensional integrated PSD. As shown in Figure \ref{fig1}(b), the parallel and perpendicular spectra at $k d_i < 0.15$ are close and slightly steeper than –1.5, consistent with previous studies \citep{cho2002compressible,zhao2022multispacecraft,hou2025energy}. At smaller scales ($k d_i > 0.15$), the perpendicular spectrum slope is nearly –3, while the parallel spectrum remains close to –1.5, indicating enhanced dissipation of quasi-perpendicular fluctuations. Despite this dissipation, the perpendicular component continues to cascade to smaller scales. After the cascade reaches ion scales ($k d_i = 1$), most of the turbulent energy concentrates on the perpendicular propagating fluctuations, while parallel propagating fluctuations decay rapidly. The presence of discrete $PSD(k_{\perp})$ peaks near ($k d_i = 1$) likely arise from fast-wave steepening, which form sharp fronts and generate small-scale waves due to dispersive effects \citep{balogh2013physics}

The electron parallel temperature ($T_{e,\parallel}$) grows most rapidly, consistent with TTD heating (Figure~\ref{fig1}c), whereas proton parallel heating ($T_{p,\parallel}$) is weaker because fewer protons satisfy the resonance condition $v_{\parallel} = \omega/k_{\parallel}$ (Figure~\ref{fig1}d) \citep{hou2025energy}. Additionally, the proton perpendicular temperature ($T_{p,\perp}$) shows a slightly stronger increase than the electron perpendicular temperature ($T_{e,\perp}$), implying an additional proton heating channel, which is attributed to resonant interactions with IBWs (Sec.~\ref{section_beam}).

\subsection{Excitation of ion Bernstein waves}

\begin{figure*}[ht]
\centering
\includegraphics[width=0.8\textwidth]{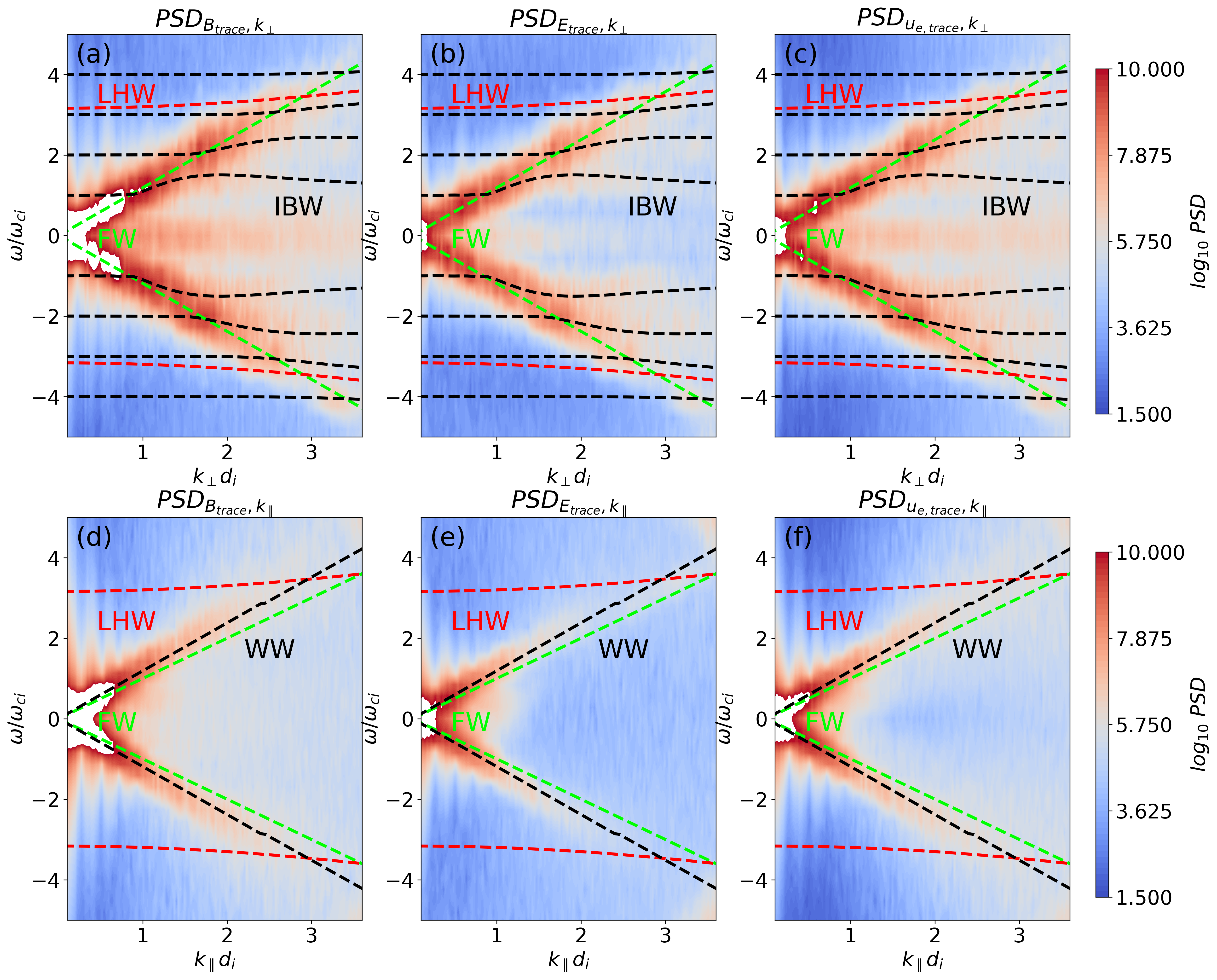}
\caption{Dispersion relations at sub-ion scales during the interval $t = 1.51\tau_A$–$1.60\tau_A$. (a)–(c) PSD Traces of the magnetic-field, electric-field, and electron-velocity fluctuation for perpendicular propagating fluctuations ($k_{\parallel} \approx 0$). The black dashed line shows the theoretical dispersion relation of the IBWs, the red line corresponds to the lower-hybrid wave (LHW), and the green line denotes the fast wave (FW) dispersion $\omega = k v_{f}$. 
(d)–(f) PSD Traces of the magnetic-field, electric-field, and electron-velocity fluctuation for parallel propagating fluctuations ($k_{\perp} \approx 0$). In panels (d)–(f), the black dashed line represents the theoretical dispersion relation of whistler waves (WW).}\label{fig3} 
\end{figure*}

IBWs at kinetic scales can be understood as a series of nearly perpendicular electrostatic eigenmodes \citep[e.g.,][]{stix1992waves,baumjohann2012basic}. They mainly propagate perpendicular to the background magnetic field, with $k_\perp \gg k_\parallel$, and are usually dominated by electrostatic fluctuations, with the perturbed electric field approximately parallel to the perpendicular wave vector. As ions gyrate around the background magnetic field, their motion produces alternating regions of particle compression and rarefaction in the plane perpendicular to the magnetic field. Thus, IBWs naturally involve density fluctuations and are compressive in nature. This density modulation leads to charge separation and builds up an electrostatic field, which provides the main restoring force and drives the plasma back toward equilibrium. Because this restoring process is modulated by finite Larmor radius (FLR) effects, the plasma can support a series of harmonic branches near the ion cyclotron frequency and its integer multiples \citep{stix1992waves,baumjohann2012basic}.

To identify waves at sub-ion scales, we perform a three-dimensional Fourier transform of the data cube $(x, y, t)$ to obtain the distribution of fluctuation energy in $\omega$–$k$ space. The Fourier analysis uses simulation data from the interval $t = 1.51\tau_A$ to $1.6\tau_A$, sampled with a high temporal resolution of $\Delta t = 0.2\Omega_p^{-1} = 1/1280\tau_A$, ensuring that the Fourier transform resolves the wave frequencies accurately.
The waves are then identified by comparing the $\omega-k$ distribution with the theoretical dispersion relations. The theoretical dispersion relations are calculated from Plasma Kinetics Unified Eigenmode Solutions (PKUES) \citep{luo2022coherence}, which is a kinetic dispersion relation solver for magnetized plasma.

Figures \ref{fig3}(a)-(c) show the PSD traces of magnetic-field, electric-field, and electron-velocity fluctuations for nearly perpendicular propagation waves ($k_{\parallel}\sim0$). Most of the fluctuation energy follows the fast wave dispersion relations, indicating the wave-like nature of compressible turbulence \citep{yuen2025temporal}. Several discrete branches are observed and are well matched by IBW dispersion relations (Figs.~\ref{fig3}(a)–(c)). For comparison, Figures \ref{fig3}(d)-(f)  display the corresponding PSD traces for nearly parallel propagation waves ($k_{\perp}\sim 0$). Here, the fluctuation energy follows the theoretical whistler wave dispersion (black dashed curve), and no discrete branches are present. This confirms that the IBWs observed in Figures \ref{fig3}(a)-(c) are not numerical artifacts. In all panels, we also include the theoretical dispersion relation, $\omega^2=k^2(T_i+T_e)/m_i+\omega_{ce}\omega_{ci}$, for the lower-hybrid wave as a reference (red dashed curve). The simulation results show no correspondence with this branch, thereby ruling out the presence of lower-hybrid waves in our simulation.

\begin{figure*}[ht]
\centering
\includegraphics[width=0.7\textwidth]{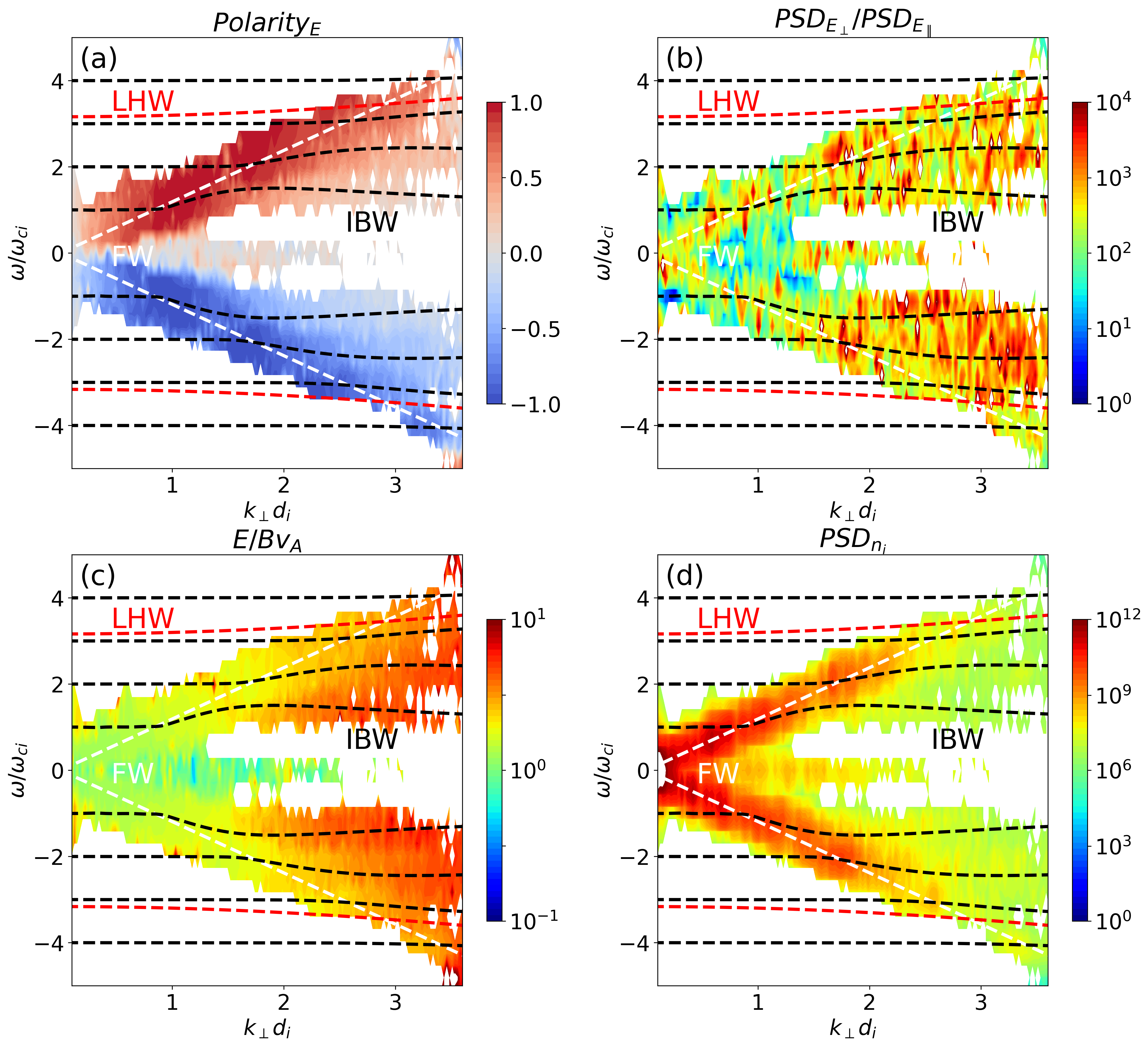}
\caption{Electrostatic nature and compressibility of perpendicularly propagating waves ($k_{\parallel}\approx0$). The black dashed line shows the theoretical dispersion relation of the IBWs, the red line corresponds to the LHW, and the green line denotes the FW dispersion $\omega = k v_{f}$
(a) Polarization state of the perpendicular electric-field components.
(b) PSD ratio of perpendicular to parallel electric-field fluctuations.
(c) Ratio of electric-field to magnetic-field.
(d) PSD ratio of proton number density.}\label{fig4} 
\end{figure*}

To further verify that the waves observed in Figures \ref{fig3}(a)–(c) are IBWs, we check the electromagnetic properties of these fluctuations. Figure \ref{fig4}(a) shows the normalized polarity of the perpendicular electric-field components, where +1 and –1 correspond to right-hand and left-hand circular polarization, and 0 corresponds to linear polarization. In the region where the fast wave energy is concentrated, the waves exhibit right-hand polarization consistent with fast-magnetosonic/whistler waves. In contrast, the waves identified as IBWs display linear polarization characteristics. Figure \ref{fig4}(b) shows the perpendicular electric field is stronger than the parallel component, indicating that the electric-field fluctuations are primarily perpendicular to B. Figure \ref{fig4}(c) shows the ratio of electric-field to magnetic-field fluctuations. This ratio greatly exceeds the Alfv\'en speed, and the electric-field fluctuations dominate the magnetic fluctuations, indicating quasi-electrostatic behavior \citep{gao2021observation}. Moreover, the waves exhibit compressibility in the proton number density (Figure \ref{fig4}(d)). Taken together, these characteristics confirm that the perpendicular propagating waves in Figures \ref{fig3}(a)–(c) are quasi-electrostatic IBWs with compressibility. Enhanced fluctuation power near the crossings of fast wave and IBW dispersion branches suggests possible mode conversion \citep{comicsel2015dispersion}, in which fast wave energy is transferred to IBWs.

\begin{figure*}[ht]
\centering
\includegraphics[width=0.6\textwidth]{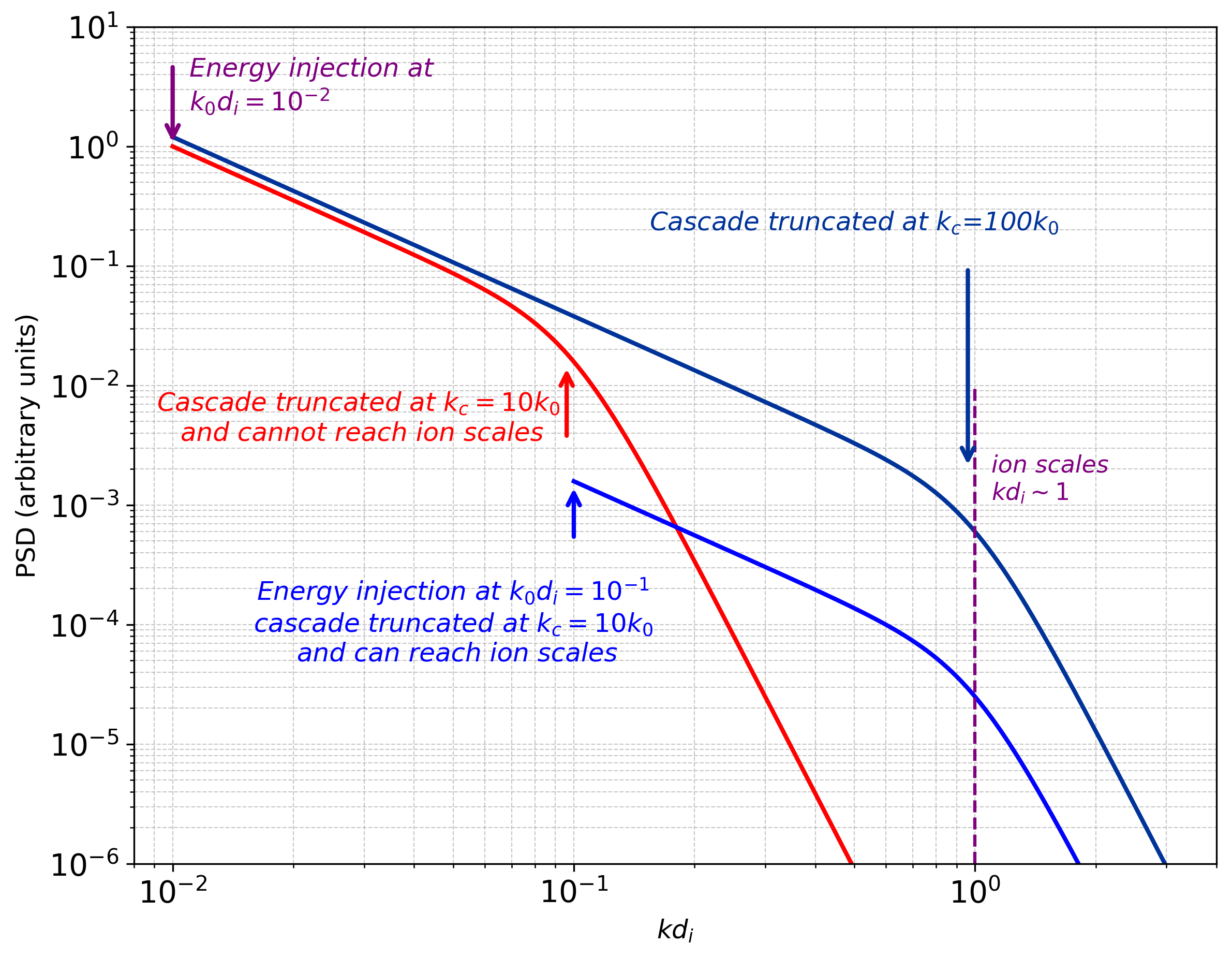}
\caption{Schematic illustration of energy-cascade behavior for different injection and truncation scales.
For injection at $k_0 d_i = 10^{-2}$ with truncation scale at $k_c = 100\,k_0$, the cascade could reach ion scales ($k d_i \sim 1$).
When the truncation scale is reduced to $k_c = 10\,k_0$, the cascade is cut off before ion scales are reached.
For a higher energy injection wavenumber ($k_0 d_i = 10^{-1}$) with the same truncation scale $k_c = 10\,k_0$, the cascade can still reach to ion scales.}\label{fig5} 
\end{figure*}

Under a realistic mass ratio ($m_p/m_e \approx 1836$) and for the plasma conditions considered here ($\beta = 0.5$), the cascade of quasi-perpendicular fast waves to ion kinetic scales is strongly suppressed by transit-time damping \citep{petrosian2006damping, yan2008cosmic}. 
However, under appropriate conditions, the turbulent cascade can reach sub-ion scales before being fully dissipated. By equaling the damping rate (Equation \ref{eq_damping}) with cascade rate ($\tau_{\mathrm{cas}}^{-1} \sim (k/L_0)^{1/2}\delta V^2/V_{\mathrm{ph}}$) \citep{cho2003compressible,yan2004cosmic}, one can obtain a $\beta$-dependent cascade truncation scale.
Here, the truncation scale refers to the scale at which the wave damping rate becomes larger than the nonlinear cascade rate, so that the turbulent cascade is effectively terminated.

When $\beta = 0.1$ and $M_A = 0.4$, the truncation scale is about one order of magnitude smaller than the injection scale, corresponding to $k_c = 10k_0$ in wavenumber space. Under lower-$\beta$ conditions, for example $\beta = 0.01$ and $M_A = 0.15$, the cascade can extend to much larger wavenumbers, with $k_c = 100k_0$. Therefore, whether the cascade can reach ion scales before being truncated depends on the plasma environment.
As shown in Figure~\ref{fig5}, when the truncation scale is $k_c = 10k_0$, fast waves need to be injected at scales corresponding to $k_0 d_i \gtrsim 0.1$ in order for the cascade to reach ion kinetic scales before truncation. Under such conditions, kinetic-scale waves can be continuously supplied by the turbulent cascade. Therefore, local compressive sources, such as magnetic reconnection accompanied by small-scale compressive waves \citep{zhuo2024oblique}, may inject fast waves and contribute to the generation of IBWs \citep{narita2016ion, renchuan2022experimental}. When the truncation scale is $k_c = 100k_0$, fast waves injected at relatively larger spatial scales ($k_0 d_i \gtrsim 0.01$) can still cascade to kinetic scales.

\subsection{Formation of particle beams}\label{section_beam}

\begin{figure*}[ht]
\centering
\includegraphics[width=1.0\textwidth]{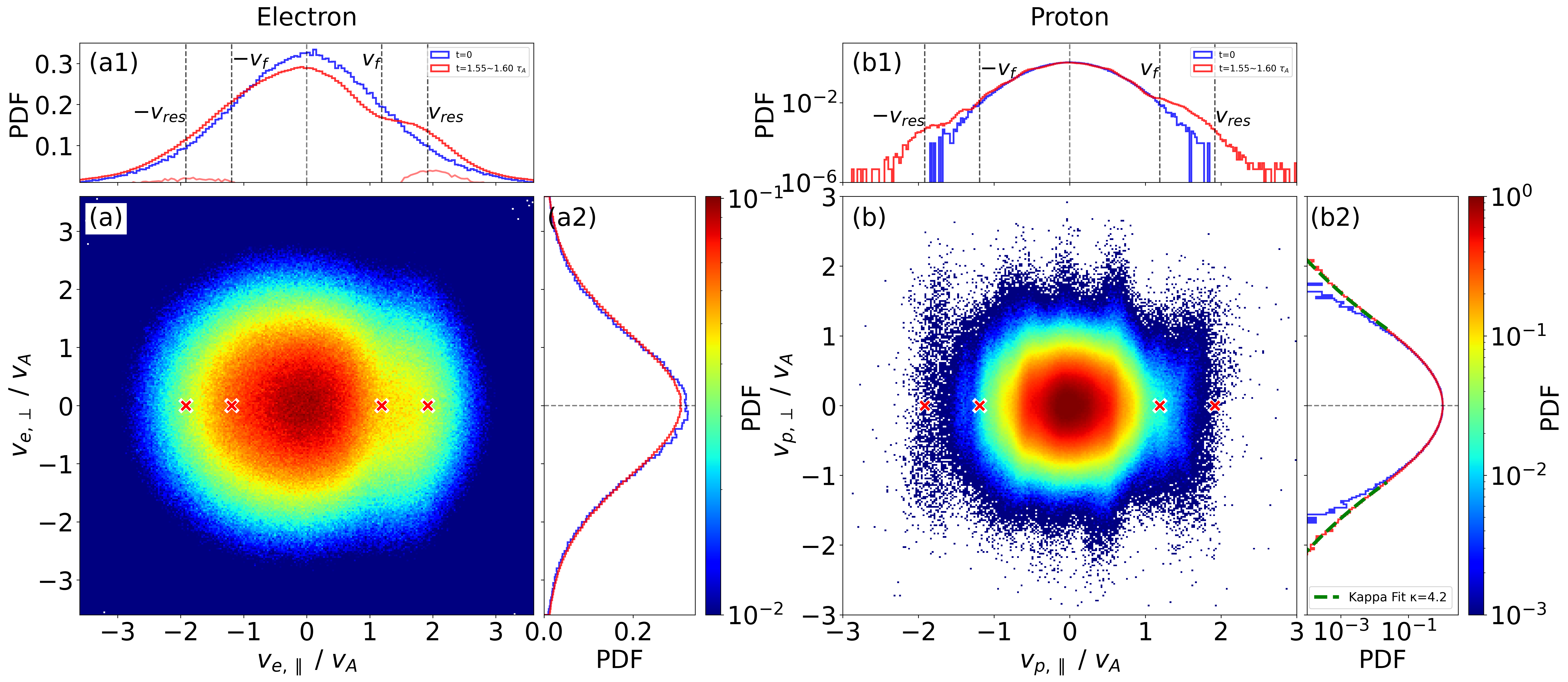}
\caption{Particle velocity distributions corresponding to the spatial region $(x = 0 $-$  2d_i, y = 0 $-$ 2d_i)$.
(a) Electron 2D velocity distribution in the $v_{\parallel}$–$v_{\perp}$ space. Red crosses mark fast magnetosonic speed $v_{f}$ and resonant speed $v_{\mathrm{res}}$.
(a1) Electron parallel velocity distribution, with dashed lines indicating $v_{f}$ and $v_{\mathrm{res}}$. 
(a2) Electron perpendicular velocity distribution. 
The blue histogram shows the initial distribution at $t=0$, and the red histogram corresponds to $t = 1.55\tau_A$–$1.6\tau_A$.
The light red lines indicate the difference between the two histograms.
(b) Proton 2D velocity distribution in the $v_{\parallel}$–$v_{\perp}$ space. Red crosses mark $v_{f}$ and $v_{\mathrm{res}}$.
(b1) Proton parallel velocity distribution, with dashed lines indicating $v_{f}$ and $v_{\mathrm{res}}$.
(b2) Proton perpendicular velocity distribution. The blue histogram shows the initial distribution at $t=0$, and the red histogram corresponds to $t = 1.55\tau_A$–$1.6\tau_A$. The green dashed lines represent a kappa distribution fit with $\kappa=4.2$.}\label{fig6} 
\end{figure*}

Figure \ref{fig6}(a) shows the velocity distribution function (VDF) of electrons in the $v_{\parallel}$–$v_{\perp}$ velocity space. A clear beam-like or non-thermal component is present, with parallel velocities exceeding the local fast magnetosonic speed $v_{f}$. The temporal evolution of the electron parallel VDF is displayed in Figure \ref{fig6}(a1), where the blue curve corresponds to the initial state at $t=0$ and the red curve to $t=1.6\tau_{A}$. Their difference indicates that the enhancement of the VDF occurs in the velocity range above $v_{f}$. This feature arises from TTD resonance between electrons and fast waves, whose condition is $v_{e,\parallel} = \omega/k_{\parallel}=\frac{v_{f}}{\cos\theta_{kB}} \ge v_{f}$, with $\theta_{kB}$ denoting the angle between the wavevector and the background magnetic field. The strongest enhancement appears near $v_{\mathrm{res}} = v_{f}/\cos45^{\circ}$. Furthermore, Figure \ref{fig6}(a2) shows that the electron perpendicular VDF remains nearly unchanged from its initial state, indicating the absence of significant perpendicular heating.

Protons exhibit similar but weaker parallel features. The proton VDF (Figure \ref{fig6}(b)) also shows a beam-like feature near $v_{f}$ and displays multiple banded structures along the perpendicular direction, indicating an increase in perpendicular temperature. Figure \ref{fig6}(b1) shows that the proton parallel VDF is enhanced at velocities greater than $v_{f}$, again due to TTD resonance with fast waves, following the condition $v_{p,\parallel} = \frac{v_{f}}{\cos\theta_{kB}} \ge v_{f}$. Since the proton thermal speed is lower than $v_{f}$, the number of protons that can satisfy the resonance condition is smaller than for electrons, resulting in weaker parallel heating. In higher-$\beta$ environments ($\beta>0.5$), more protons are expected to participate in TTD resonance and resonate with a larger $v_{f}$. In higher-$\beta$ plasmas, a larger fraction of protons is expected to resonate with fast waves, leading to increased proton beam density and drift speed. 
Unlike electrons, protons exhibit clear perpendicular energization (Figure~\ref{fig6}(b)), with velocity distributions displaying kappa-like suprathermal tails with $\kappa=4.2$ (Figure~\ref{fig6}(b2)). Notably, enhanced perpendicular velocities are observed even for protons with $v_{p,\parallel}\approx 0$, indicating that this heating is unlikely to result from cyclotron resonance with right-hand polarized whistler waves, which requires $v_{p,\parallel} > v_f$. Instead, the observed behavior is more naturally explained by resonant interactions with IBWs \citep{narita2016ion}.

Under a realistic mass ratio $m_p/m_e \approx 1836$, the TTD damping rate (Equation~\ref{eq_damping}) predicts that protons predominantly resonate with fast waves propagating at angles $\theta_{kB} \sim 30^\circ$--$50^\circ$ relative to the background magnetic field. The corresponding proton resonant velocity depends on $\beta$. 
For $\beta = 0.5$, the proton-dominated maximum TTD damping rate occurs at a propagation angle of $\theta_{kB} \approx 34^\circ$ (Figure \ref{fig2}), corresponding to a resonant velocity of $v_{\rm res} \approx 1.3\,V_A$. This naturally produces a proton beam with the same drift speed ($1.3\,V_A$). As $\beta$ increases, a larger fraction of protons satisfies the resonance condition, leading to a substantial enhancement of the TTD damping rate (Figure \ref{fig2}). When $\beta = 2$, the maximum damping shifts toward more oblique propagation, occurring at $\theta_{kB} \approx 50^\circ$, while the corresponding resonant velocity increases to $v_{\rm res} \approx 2.4\,V_A$, well into the super-Alfv\'enic regime. Such super-Alfv\'enic drift speeds are consistent with in-situ measurements by PSP near the Sun \citep{verniero2022strong}.

When proton moments are computed without separating core and beam, solar wind shows an inverse dependence between the proton temperature anisotropy ($T_{p,\perp}/T_{p,\parallel}$) and $\beta$ \citep{huang2020proton, huang2025temperature}. In addition, in the slow solar wind, which contains a more compressive component, a greater fraction of measurements exhibits higher $T_{p,\parallel}$ \citep{huang2020proton, huang2025temperature}. These observations may arise from multiple physical processes. Nevertheless, they are consistent with our expectations according to the TTD process. Consequently, the super-Alfv\'enic proton beams in the solar wind can originate, at least in part, from the damping of fast waves via the TTD mechanism.

\section{Conclusion}

To investigate the source of particle beams and IBWs, we perform numerical simulations of fast wave turbulence from MHD to sub-ion scales using a reduced ion–electron mass ratio. The principal findings are summarized as follows.
\begin{itemize}

    \item At MHD scales, TTD of compressible turbulence can drive the formation of particle beams. The proton resonant speed increases with plasma $\beta$, allowing the resulting beam drift to become strongly super-Alfvénic under typical solar wind conditions, consistent with near-Sun PSP observations.

    \item Under low-$\beta$ conditions, IBWs are generated as confirmed by their dispersion relations, quasi-electrostatic nature, and compressibility. IBWs could emerge from intrinsic coupling with fast waves, facilitated by compressible turbulent evolution. IBWs offer an efficient channel for perpendicular energization of suprathermal protons. At higher $\beta$ ($\beta>1$), strong TTD suppresses the turbulent cascade \citep{petrosian2006damping,yan2008cosmic}, while enhanced TTD-generated proton beams may provide an additional pathway for IBW excitation through kinetic instabilities.

\end{itemize}

\begin{acknowledgments}
We acknowledge the computing resources from the high-performance computers at the NHR center NHR@ZIB, with the project No. bbp00080.
The VPIC simulation code is available at \url{https://github.com/lanl/vpic}. We would like to acknowledge the use of ChatGPT for improving the English grammar and sentence structure. C.H. is supported by the Alexander von Humboldt Foundation.
\end{acknowledgments}

\end{document}